\documentclass[12pt]{article}

\usepackage[a4paper,margin=1in]{geometry}
\usepackage[T1]{fontenc}
\usepackage[utf8]{inputenc}
\usepackage{lmodern}
\usepackage{setspace}
\usepackage{titlesec}
\usepackage{hyperref}
\hypersetup{hidelinks}

\title{Henri Poincaré's Saint Louis Lecture of 1904: Publication, Dissemination, and Historiographical Implications}

\author{Hector Giacomini\\[0.4em]
\small Institut Denis Poisson. Universit\'e d'Orl\'eans --- Universit\'e de Tours --- CNRS.\\
\small 37200 Tours, France\\
\small \texttt{giacominihector@gmail.com}}

\date{}

\makeatletter
\renewcommand{\maketitle}{%
  \begin{center}
    {\LARGE \@title \par}
    \vspace{1em}
    {\large \@author \par}
  \end{center}
  \vspace{0.8em}
}
\makeatother

\begin{document}

\maketitle

\begin{abstract}
\small
Henri Poincaré's Saint Louis lecture of 1904 occupies an important place in the prehistory of relativity. In it, Poincaré formulated the principle of relativity in general terms and presented it as one of the guiding principles of mathematical physics, together with the principles of least action and energy conservation. This article reconstructs the early publication and international dissemination of the lecture before the end of June 1905, through \textit{La Revue des idées}, the \textit{Bulletin des sciences mathématiques}, \textit{The Monist}, and \textit{La valeur de la science}. Drawing on library records, accession data, booksellers' advertisements, press notices, and correspondence, it shows that Poincaré's text circulated rapidly through scholarly, commercial, and institutional channels in Europe and North America.

The significance of this circulation is not merely bibliographical. It bears directly on the documentary landscape within which Einstein's 1905 relativity paper should be historically situated. The early availability of Poincaré's lecture and of \textit{La valeur de la science} shifts attention toward the concrete conditions under which texts, concepts, and problems circulated in the months and weeks preceding Einstein's June 1905 paper. Evidence from Einstein's Bern milieu further supports this view. Joseph Sauter's later testimony and the possibility of redating the Habicht letter to 1 June 1905, on the basis of several converging chronological and documentary arguments, suggest that the intellectual environment of 1905 was denser than simplified narratives of solitary discovery imply. Availability is not influence; but without reconstructing availability, the historical problem of Einstein's relation to Poincaré and Lorentz remains ill posed.
\end{abstract}

\section{Introduction}

Poincaré's Saint Louis lecture has long been recognized as an important document in the history of the principle of relativity. Yet its historical significance does not depend only on the conceptual content of the lecture itself. It also depends on the forms in which the text was printed, reprinted, translated, advertised, catalogued, acquired, and made available to readers in the months immediately preceding Einstein's 1905 paper on the electrodynamics of moving bodies. The present article approaches the Saint Louis lecture from this material and documentary perspective.

The question is not simply whether Poincaré formulated ideas that later became relevant to the history of relativity. That point has often been discussed in the historiography. The more elementary question, prior to any claim about influence, is whether the relevant texts were in circulation early enough, and through sufficiently ordinary scholarly channels, to form part of the documentary environment of 1904--1905. In this sense, the article does not begin from a hypothesis of direct reading by Einstein. It begins from a more limited historical task: to reconstruct the actual conditions of availability under which any such question must be assessed.

The inquiry therefore follows the successive printed forms of the Saint Louis lecture: its appearance in \textit{La Revue des idées}, its publication in the \textit{Bulletin des sciences mathématiques}, its English version in \textit{The Monist}, and its incorporation into \textit{La valeur de la science}. These publication channels were not equivalent. They addressed different publics, moved through different networks, and left different kinds of documentary traces. Taken together, however, they show how a lecture delivered in September 1904 entered several overlapping circuits of intellectual, mathematical, philosophical, and institutional communication before the summer of 1905.

This reconstruction also explains why the final part of the article turns to Einstein's Bern milieu. The point is not to displace the focus from Poincaré's lecture to a speculative account of Einstein's sources, but to clarify why the chronology of circulation matters. If \textit{La valeur de la science} was already being advertised, acquired, and read in April and May 1905, then the relevant historical question cannot be framed as though Poincaré's text belonged only to a later reception history. The problem becomes instead one of documentary proximity: which texts were available, where, through which channels, and within what institutional and social environment?

Several pieces of evidence are considered in this connection: the early circulation of \textit{La valeur de la science}, Maurice Solovine's 1952 report to Carl Seelig, who was then preparing a biography of Einstein, that the readings of the Olympia Academy included both \textit{La science et l'hypothèse} and \textit{La valeur de la science}, the chronology of Einstein's letter to Conrad Habicht announcing the four papers of 1905, and Joseph Sauter's later recollection of Einstein's discussions at the Bern Patent Office. None of these materials proves that Einstein read Poincaré's Saint Louis lecture before writing his relativity paper. They do, however, make it necessary to situate Einstein's work within a dense documentary environment rather than within a simplified narrative of intellectual isolation.

The proposed redating of the Habicht letter is especially relevant here. The traditional dating to 18 or 25 May 1905 leaves open a narrow interval between the completion of Einstein's other papers and the final drafting of the relativity paper. A date of 1 June 1905, while not demonstrable with certainty, fits several chronological indications and would place the letter still closer to the final phase of Einstein's work. This possibility matters because it enlarges the immediate documentary window in which recently published material, including \textit{La valeur de la science}, may have been accessible in Bern.

The Bern setting is therefore treated as part of the material history of scientific communication. Einstein's access to the University Library of Bern, his reviewing activity for the \textit{Beiblätter zu den Annalen der Physik}, the availability of major contemporary physics periodicals, and the circulation of international scientific books all complicate the image of a young physicist working in documentary isolation. The article argues that the history of availability is the necessary ground on which more precise questions about influence, silence, and retrospective memory must be posed.

\section{Scientific Publication, International Exhibitions, and the Circulation of Poincaré's Lecture}

The publication history of Poincaré's Saint Louis lecture should be understood as part of the infrastructure through which scientific authority, priority, visibility, and international circulation were established around 1900. Recent historiography has emphasized that journals and related publication formats helped shape the public life of scientific knowledge. Alex Csiszar \cite{Csiszar2018} has shown how the scientific journal became, during the nineteenth century, a central institution for authorship, credibility, judgment, and the politics of scientific communication. Melinda Baldwin \cite{Baldwin2014, Baldwin2015} has likewise drawn attention to the importance of publication speed, editorial strategy, and international competition in the scientific periodical press, especially in relation to \textit{Nature} and to the changing norms of physics publishing at the turn of the twentieth century. Poincaré's Saint Louis lecture offers a revealing case of this broader publication culture, because the same address circulated within a few months through several distinct formats: a French intellectual review, a mathematical periodical, an American philosophical and scientific journal, and a widely distributed book.

This perspective is directly relevant to the Saint Louis lecture. Poincaré's address was delivered at the International Congress of Arts and Science, held as part of the Louisiana Purchase Exposition of 1904. The official report of the congress \cite{PoincareS} presented it as an ambitious attempt to organize and display the state of knowledge across disciplinary boundaries. The setting itself matters. International exhibitions and world fairs were major sites for the staging of progress, national prestige, imperial competition, industrial modernity, and the public circulation of scientific and technical culture. The Saint Louis Congress must therefore be situated within the broader culture of international exhibitions, whose official records and historical analyses show how such events functioned as public stages for scientific exchange, national representation, and the international circulation of knowledge \cite{Rogers1905,Rydell1984,Greenhalgh1988}. The Paris Exposition of 1900 had already presented science, electricity, instruments, industry, and international exchange as central features of modern civilization. As Richard Staley has shown, such exhibitions created a cultural and institutional space in which physics, measurement, instrumentation, and questions about the foundations of modern science were displayed, compared, and publicly discussed \cite{Staley2008}. The Saint Louis Congress belongs to this broader sequence of international exhibitions and congresses, in which scientific lectures were staged as public interventions and then extended through print.
This context helps explain why the subsequent publication and republication of Poincaré's address mattered. A lecture delivered in such a setting could acquire a life beyond the moment of oral presentation through printed versions, translations, journal appearances, book republication, library accessions, and commercial announcements. These forms belonged to the same international machinery through which scientific ideas acquired visibility and mobility. To reconstruct the circulation of the Saint Louis lecture is therefore to recover the material conditions under which a major intervention on the principles of mathematical physics became available to readers in different national and institutional settings before Einstein submitted his 1905 relativity paper.

Seen in this context, the publication of Poincaré's lecture in several formats is historically significant. The lecture moved from an international congress to French, American, and book channels: it appeared in \textit{La Revue des idées}, in the \textit{Bulletin des sciences mathématiques}, in \textit{The Monist}, and then in \textit{La valeur de la science}. These supports were not equivalent. They belonged to different intellectual and commercial circuits, addressed different readerships, and carried different implications for the speed and geography of circulation. The same text could therefore reach several partially overlapping publics: philosophical and scientific readers, mathematicians and physicists following the \textit{Bulletin des sciences mathématiques}, readers of the transatlantic forum represented by \textit{The Monist}, and the broader educated readership reached by Poincaré's book.

This matters for the historiography of relativity. The question is not simply whether Poincaré's Saint Louis lecture was printed before Einstein submitted his 1905 relativity paper, but how, where, in what forms, and through which institutional and commercial channels it became available before Einstein submitted his relativity paper at the end of June 1905. The dates of publication, reception, acquisition, and advertisement examined in this article reconstruct the documentary landscape in which a major text by Poincaré on principles, relativity, mechanics, and mathematical physics circulated before Einstein's paper was submitted.

The argument developed here concerns the conditions of historical interpretation. Establishing the material circulation of Poincaré's lecture changes the terms in which the question of Einstein's possible knowledge of it should be posed. Once rapid and plural circulation has been documented, explanations based on late, narrow, or inaccessible publication lose much of their force. The historical problem then shifts from simple availability to more precise questions of readership, intellectual assimilation, and citation practice.

\subsection{Early Notices in German Mathematical Periodicals}

An early indication of the visibility of the Saint Louis Congress in the European scientific community is provided by contemporary German mathematical periodicals. In June 1904, the \textit{Jahresbericht der Deutschen Mathematiker-Vereinigung}, the principal journal of the German Mathematical Society, published a notice announcing the forthcoming International Congress of Arts and Science in Saint Louis. The notice explicitly mentioned several of the principal lecturers associated with the mathematical sections of the congress, including Henri Poincaré, Ludwig Boltzmann, Émile Picard, Gaston Darboux, Edward Kasner, and Heinrich Maschke. The presence of these names in a German mathematical journal several months before the congress shows that the event was already known and recognized within the European mathematical community.

A further sign of the lecture's rapid reception can be found in Göttingen. On 31 January 1905, at the \textit{Mathematische Gesellschaft}, C.~H.~Müller presented, at Felix Klein's request, a report on the mathematical communications delivered at the Saint Louis Congress. He devoted particular attention to Poincaré's address \cite{JDMV1905}. This is an important institutional witness. Göttingen was one of the principal European centres of mathematical and physical research, and Klein's circle was deeply engaged with the relation between mathematics, mechanics, physics, and foundations. The discussion of Poincaré's lecture there in January 1905 shows that the Saint Louis address had already entered organized scholarly discussion in Germany several months before Einstein submitted his relativity paper.

These early German notices therefore complement the later publication and accession evidence examined below. They show that the Saint Louis Congress and Poincaré's contribution to it were objects of attention within European mathematical networks by early 1905.

\section{The Scientific Content of the Saint Louis Lecture}

Poincaré's Saint Louis lecture of 24 September 1904 may be read as a synthetic assessment of the state of mathematical physics rather than as a technical research contribution. It offered a broad diagnosis of what Poincaré described as a possible ``crisis'' in the foundations of physics, touching upon mechanics, thermodynamics, electrodynamics, and the theory of electrons.

A central theme of the lecture was the status of fundamental principles in theoretical physics. Poincaré explicitly enumerated the guiding principles of contemporary theory: the conservation of energy, or principle of Mayer; the principle of Carnot, or degradation of energy; the equality of action and reaction, or principle of Newton; the conservation of mass, or principle of Lavoisier; the principle of least action; and, crucially, the principle of relativity. These principles were not presented as consequences of detailed mechanical hypotheses, but as general constraints capable of guiding theory even when the microscopic mechanisms of nature remained uncertain.

Poincaré contrasted an earlier physics grounded in atomistic models and central forces with a more recent ``physics of principles,'' in which emphasis shifts from mechanical construction to general constraints on theory. Principles originally abstracted from experience acquire methodological primacy: they delimit admissible theoretical constructions and provide coherence across otherwise distinct physical domains.

\subsubsection*{The Principle of Relativity}

Poincaré formulated the principle of relativity in explicit and programmatic terms: the laws of physical phenomena must be the same for an observer at rest and for an observer carried along in uniform translational motion. Consequently, no experiment should allow one to determine whether one is in such uniform motion or not. In the lecture this formulation appears within the canonical list of fundamental principles and is treated on the same conceptual level as energy conservation and the principle of least action.

The relativity principle is not introduced merely as an empirical summary of ether-drift experiments. Poincaré presents it as a general requirement increasingly supported by the persistent failure of attempts to detect motion relative to the ether. He discusses in particular the negative results of Michelson-type experiments and emphasizes the remarkable stability of electromagnetic theory under uniform motion. The repeated confirmation of null results is interpreted as evidence that the invariance of physical laws under uniform translation may express a deep property of nature.

Some commentators have noted that Poincaré's formulation refers explicitly to observers, whereas later formulations speak directly of the invariance of the laws of nature. Taken in isolation, this observer-language might seem compatible with a privileged frame in which the laws have a simpler form. The broader context of the lecture weakens such a reading. Poincaré's formulation concerns the \textit{laws of phenomena}, not merely the limitations of observation, and the failure of ether-drift experiments is presented as pointing toward a general invariance rather than a merely perceptual relativity. The emphasis therefore falls on the structural stability of the laws governing physical processes under uniform translation.

In this respect, Poincaré's principle of relativity appears as the explicit crystallization of themes already articulated in his earlier writings \cite{Poincare1899, Poincare1900, Poincare1902}, and made accessible to German-speaking readers through the 1904 translation of \textit{La science et l'hypothèse} \cite{Poincare1904}. He further analyzes the theoretical devices introduced to preserve this invariance. He discusses Lorentz's notion of ``local time,'' obtained by synchronizing clocks through light signals under the assumption of equal light speeds in opposite directions, and notes that observers in uniform motion would naturally adopt such a synchronization without detecting any measurable effect revealing their motion. He also refers to length contraction in the direction of motion and to the modifications of dynamical quantities required to maintain agreement with experiment. These elements appear as components of an increasingly elaborate theoretical effort to safeguard the relativity principle.

Importantly, Poincaré suggests that the situation may ultimately require a new mechanics in which no velocity could exceed that of light and in which inertia would increase with speed. The principle of relativity is therefore not treated as a peripheral correction within classical mechanics, but as a constraint capable of reshaping its conceptual structure.

In this respect, the Saint Louis lecture belongs to Poincaré's broader conventionalism: the principle of relativity is not presented simply as an empirical regularity, but as an organizing principle whose status depends on its capacity to preserve the coherence of physical theory in the face of experimental and conceptual difficulties.

\subsubsection*{Lorentz's 1904 Theory}

An important component of the lecture is Poincaré's discussion of Hendrik Antoon Lorentz's recent work. In 1904 Lorentz had published a major paper presenting a refined mathematical formulation of the transformations required to preserve the form of Maxwell's equations in a moving frame \cite{Lorentz1904}.

Poincaré's lecture shows that by September 1904 he was closely aware of the main elements and significance of Lorentz's construction. He describes the introduction of local time, the contraction hypothesis according to which bodies moving through the ether undergo a physical contraction in the direction of motion, and the modification of forces and masses required to reconcile theory with experiment. In Lorentz's framework, this contraction was introduced as a specific dynamical hypothesis intended to account for the null results of ether-drift experiments, most notably the Michelson--Morley experiment. For Poincaré, the growing complexity of these hypotheses reflected the seriousness of the challenge posed by the relativity principle and by the experimental results of Michelson and others.

Although Lorentz maintained the ether as a privileged frame and introduced the transformed time $t'$, which he called ``local time'' and associated with the time indicated by clocks in the moving system, Poincaré's interpretation shifts the emphasis. The preservation of the form of Maxwell's equations under transformation is presented not simply as a technical success, but as evidence of an underlying invariance deserving elevation to principled status.

\subsubsection*{From Technical Devices to Structural Invariance}

While recognizing the mathematical ingenuity of Lorentz's theory, Poincaré places the interpretative focus on invariance itself. The stability of electromagnetic laws under uniform translation is portrayed as a feature that theoretical physics must take seriously. Rather than grounding theory in mechanical models of the ether, he advocates an approach guided by general principles whose empirical resilience suggests deep validity.

In this framework, the transformations are not merely compensatory devices designed to preserve an ether theory. They signal the possibility that classical mechanics itself may require modification. Poincaré explicitly contemplates a future mechanics characterized by velocity-dependent inertia and by the impossibility of exceeding the speed of light. The lecture thus articulates a conception of physics centered on symmetry, invariance, and coherence between domains, rather than on detailed mechanical modeling.

The rapid publication of this lecture in several venues between November 1904 and January 1905 therefore ensured that these reflections became available for international scientific discussion almost immediately.

\section{Historiographical Assessments}

A substantial body of scholarship has recognized the importance of Poincaré's Saint Louis lecture, although the precise status attributed to the text has varied. Historians and philosophers of physics generally agree that the lecture marks a moment of unusual explicitness in Poincaré's treatment of the relativity principle and that it occupies a significant place in the sequence linking late nineteenth-century electrodynamics to the conceptual reconfigurations of the early twentieth century.

Several authors have emphasized the principled character of Poincaré's formulation. Michel Paty \cite{Paty1992, Paty1996, Paty1998} stressed that Poincaré did not merely invoke the principle of relativity by name but assigned to it a universal scope and a methodological role in the organization of theoretical physics. In Paty's interpretation, mathematical physics is structured around a limited set of general principles---including the equality of action and reaction, the conservation of mass and energy, the second law of thermodynamics, the principle of least action, and the principle of relativity---which function as guiding constraints in the construction of physical theories. Within this framework, the relativity principle appears not as a hypothesis restricted to electrodynamics but as a foundational requirement governing the mathematical representation of physical laws. Raffaella Toncelli \cite{Toncelli2013} likewise interprets Poincaré's relativity through the architectonic function of general principles, while Roberto Torretti \cite{Torretti1983} emphasized the programmatic character of Poincaré's formulation. Elie Zahar \cite{Zahar1989} reproduced passages from the Saint Louis lecture in his discussion of the transition from Lorentzian electrodynamics to later theoretical developments. More recently, Jean-Marc Ginoux \cite{Ginoux2024} has underlined the importance of the lecture, emphasizing Poincaré's explicit formulation of the relativity principle, his discussion of ether-drift experiments, and his anticipation of a new mechanics in which the velocity of light would constitute an upper limit.

A second line of interpretation has treated the Saint Louis lecture as a major document for understanding the conceptual state of physics immediately before 1905. In an early study, I.~Yu.~Kobzarev \cite{Kobzarev1974} read the text as a revealing synthesis of the tensions affecting theoretical physics at that moment, including the interpretation of Maxwell's equations, the status of the ether, the statistical interpretation of thermodynamics, and the emerging formulation of the relativity principle. Giuliano Giannetto \cite{Giannetto1995} developed this analysis in greater detail, interpreting the lecture not merely as a survey of contemporary difficulties, but as a carefully organized diagnosis of the crisis of classical mechanics and of the need for a new theoretical framework.

Olivier Darrigol \cite{Darrigol2012, Darrigol2022} has likewise situated the lecture within Poincaré's broader analysis of the crisis of fundamental principles in early twentieth-century physics. In his reading, the lecture integrates the most recent developments of Lorentz's 1904 electron theory into a systematic diagnosis of the difficulties affecting the major principles of physics. Carnot's principle appears threatened by the statistical interpretation of irreversibility and by Brownian motion; the principle of reaction is challenged by radiation pressure and electromagnetic momentum; and the principle of relativity itself seems endangered within a theory still based on a stationary ether. At the same time, Darrigol emphasizes that the lecture contains one of Poincaré's clearest early formulations of the relativity principle, defined as the requirement that the laws of physical phenomena be the same for an observer at rest and for an observer carried along in uniform translational motion.

Richard Staley \cite{Staley2008} has given particular weight to the Saint Louis lecture in his reconstruction of the cultural and intellectual setting of relativity. In his account, before Einstein's 1905 paper Poincaré had gone further than any other physicist in bringing together the search for new electrodynamic foundations and the philosophical critique of the central concepts of mechanics. Staley emphasizes that, already in \textit{La science et l'hypothèse}, Poincaré had questioned the absolute status of space and time and had treated simultaneity as a concept requiring critical analysis. The Saint Louis lecture then appears as a further stage in this trajectory: Poincaré placed the principle of relativity among the endangered principles of mathematical physics, connected it with the difficulties of Lorentzian electrodynamics, and suggested the possibility of a new mechanics in which inertia would increase with velocity and the velocity of light would become an unsurpassable limit. Staley nevertheless regards the lecture itself as a text that Einstein did not know, while recognizing that Einstein had read \textit{Wissenschaft und Hypothese} avidly and that Poincaré's philosophical criticism formed part of the broader intellectual background against which Einstein reworked the relation between mechanics and electrodynamics. This assessment is especially important for the present article, since it identifies the Saint Louis lecture as a text of major conceptual significance while leaving its material availability to Einstein outside the analysis.

Other historians have focused on more specific conceptual aspects. Scott Walter \cite{Walter2014} has drawn attention to Poincaré's discussion of clock synchronization and to his analysis of a possible empirical challenge to Lorentz's theory arising from hypothetical signals propagating faster than light. Such signals, Poincaré observed, could in principle reveal discrepancies in the synchronization of clocks adjusted by light signals and thereby disclose the common motion of the clocks, contradicting the relativity principle. Arthur I.~Miller \cite{Miller1998} likewise noted that in the Saint Louis lecture Poincaré presented Lorentz's theorem of corresponding states as an expression of the relativity principle, while Stanley Goldberg \cite{Goldberg1984} emphasized the explicit and general character of Poincaré's statement that the laws of physical phenomena must be the same for observers in uniform translational motion.

This later citation history confirms the need to distinguish material availability before Einstein's June 1905 relativity paper from subsequent patterns of recognition: as Gingras has shown, Poincaré's work was only weakly connected to Einstein's in early relativity citation practices, while the stronger Einstein--Poincaré association emerged largely in later historiographical memory \cite{Gingras2008}.

This historiographical recognition makes the question of circulation all the more significant. If the Saint Louis lecture represented a major moment in the explicit formulation of the relativity principle, then its publication between late 1904 and early 1905 disseminated not a peripheral technical remark but a text situated at the highest level of contemporary theoretical reflection. 

\section{The First Printed Publication: \textit{La Revue des idées} (15 November 1904)}

The first printed appearance of Poincaré's Saint Louis lecture occurred not in a specialized scientific journal, but in the Parisian intellectual review \textit{La Revue des idées}, in the issue dated 15 November 1904 \cite{PoincareL}. Founded earlier that year, the journal was a monthly periodical, appearing on the fifteenth of each month, and addressed to a cultivated readership interested in philosophy, science, literature, and contemporary intellectual debate.

Unlike learned-society journals, \textit{La Revue des idées} operated within a commercial publishing framework. A contemporary notice in the \textit{Revue de linguistique et de philologie comparée} announced the inaugural issue in explicit terms: ``Le numéro 1 de la \textit{Revue des Idées} (1 fr.~50, chez tous les libraires) vient de paraître.'' This statement shows that, from the outset, the review was marketed through the general retail book trade in France and was available for individual purchase beyond subscription alone. The same notice also highlighted the scientific contents of the first issue, indicating that science formed part of the journal's public identity from its launch.

Its intellectual profile was explicitly interdisciplinary. Tables of contents from 1904 show contributions in philosophy, mathematics, physics, biology, history, and literary studies, often written by university professors or members of major French scholarly institutions. Within this editorial environment, Poincaré's article, ``L'État actuel et l'avenir de la physique mathématique,'' appeared as part of a broader intellectual discussion rather than as a technical communication restricted to specialists. The publication therefore inserted Poincaré's formulation of the relativity principle into the commercially organized culture of Belle Époque intellectual periodicals.

Evidence of active circulation appeared almost immediately. The inaugural issue of \textit{La Revue des idées} was described in detail in the Brussels weekly \textit{L'Art moderne} of 14 February 1904, less than one month after the publication of the first issue. This early notice is significant because it shows that the new Parisian review was already being monitored outside France at the moment of its launch. The journal was not confined to a narrow Parisian readership; it entered a wider francophone and European culture of periodical exchange in which intellectual reviews were read, summarized, and evaluated across national borders.

The strongest evidence for the rapid visibility of the Poincaré issue itself comes from \textit{L'Enseignement mathématique}. Founded in 1899 and edited from Geneva and Paris, \textit{L'Enseignement mathématique} was not simply a pedagogical journal. It also functioned as a transnational forum for broader questions concerning the foundations, meaning, and development of mathematics and mathematical physics. Its prestige is reflected in the presence, among its contributors during the first decade of the twentieth century, of figures such as Henri Poincaré, Paul Appell, Paul Painlevé, Gaston Darboux, Émile Borel, Felix Klein, Guido Castelnuovo, and Adolf Hurwitz.

A revealing episode appears in this context. In vol.~6 (1904), \textit{L'Enseignement mathématique} published ``Définition physique de la force,'' a communication presented at the Second International Congress of Philosophy in Geneva in September 1904 by Louis Hartmann. In vol.~7 (January 1905), p.~41, the journal printed a brief letter from Ernst Mach, who found Hartmann's ideas natural and interesting from the standpoint of the historical development of mechanics. The editors then added a remark from an anonymous reader suggesting that Hartmann's communication be compared with Henri Poincaré's Saint Louis lecture on the present state and future of mathematical physics.

The editors did not merely relay this comparison. They supplied the bibliographical information themselves, noting that Poincaré's lecture had been reproduced in \textit{La Revue des Idées} of 15 November 1904. The chronology is striking: barely a month and a half after the publication of that issue, an explicit reference to it appeared in \textit{L'Enseignement mathématique}. This editorial gesture shows that Poincaré's lecture had already become visible enough to be invoked in an ongoing discussion on force, motion, and the principles of mathematical physics. It is therefore one of the clearest contemporary witnesses to the rapid scholarly circulation of the first printed version of the Saint Louis lecture.

Surviving library holdings confirm the subsequent integration of \textit{La Revue des idées} into major academic collections across Europe and North America. Complete or substantial runs are preserved at the Bibliothèque de Genève, the Bibliothèque universitaire de Genève, and the Bibliothèque cantonale et universitaire de Fribourg in Switzerland; the University of Vienna; the Staatsbibliothek zu Berlin; the Universitätsbibliothek Erlangen--Nürnberg; the Koninklijke Bibliotheek in The Hague; the Royal Danish Library in Copenhagen; the Biblioteca Nazionale Centrale di Roma; the British Library in London; the Royal Library of Belgium in Brussels; and several major North American research libraries, including Princeton University, Yale University, Harvard University, the University of Chicago, Stanford University, the University of Illinois Urbana--Champaign, and the Library of Congress. These holdings should be read together with the contemporary notices and references discussed above. They show that a review whose first numbers were already commercially advertised and rapidly monitored entered durable scholarly collections on both sides of the Atlantic.

The available archival documentation indicates that the university and research-library holdings mentioned above were not necessarily acquired at the moment of publication. Several institutional collections were assembled retrospectively through the international book trade. Commercial intermediaries such as the Brussels bookseller Falk, the Viennese firm Carl Gerold and Co., the Turin-based bookseller Ermanno Loescher, Hubert Welter in Paris, and the New York importer G.~E.~Stechert and Company supplied European periodicals to research libraries as commission agents, subscription managers, and retail booksellers integrated into transnational distribution networks. Their documented role confirms the commercial channels through which a periodical such as \textit{La Revue des idées} could circulate beyond its immediate Parisian setting. The journal circulated through subscription, through retail sale in French bookshops, and through the wider book-trade mechanisms that connected Parisian periodical publishing to international scholarly libraries.

Taken together, the evidence shows that Poincaré's Saint Louis lecture first appeared in a commercially distributed intellectual review whose scientific content was visible from its launch and whose Poincaré issue was explicitly cited in a major mathematical periodical by January 1905. The first printed publication of the lecture therefore already placed the text within a network of intellectual review culture, mathematical discussion, retail book trade, and academic collection several months before Einstein submitted his relativity paper.

\section{\textit{The Monist} (1890--1905): A Transatlantic Forum for Science, Philosophy, and Foundations}

Founded in 1890 by Paul Carus and published by the Open Court Publishing Company, \textit{The Monist} occupied a distinctive position in the intellectual landscape of the fin de siècle. Although often described as a philosophical periodical, its contents between 1890 and 1905 reveal a broader profile. Alongside discussions of metaphysics, religion, and the unity of knowledge, the journal regularly published essays on mathematical physics, geometry, logic, probability, atomism, energetics, physical space, and the conceptual foundations of the exact sciences. It therefore functioned as a transatlantic forum in which philosophical reflection and scientific analysis could intersect.

This profile is essential for understanding the significance of the English publication of Poincaré's lecture there in January 1905 \cite{PoincareT}. \textit{The Monist} was not an accidental or marginal venue for a lecture on the foundations of mathematical physics. It had already become a recognizable outlet for authors concerned with the structure of scientific knowledge, the status of mathematical reasoning, and the interpretation of modern physical theory. The publication of the Saint Louis lecture in this journal gave Poincaré's text an English-language and transatlantic form of rapid circulation.

\subsection{Major Figures and Foundational Questions}

Before 1905, \textit{The Monist} had published contributions by several major figures in mathematics, physics, logic, and epistemology. Poincaré himself had already appeared in its pages with two substantial essays: ``On the Foundations of Geometry'' in 1898 and ``Relations between Experimental and Mathematical Physics'' in 1902. These texts examined the logical structure of geometry and the relation between theory and experiment, and they prepared the readership of \textit{The Monist} for the kind of foundational reflection developed in the Saint Louis lecture.

The same can be said of other prominent contributors. Ernst Mach used \textit{The Monist} as a recurrent outlet for his empirically oriented philosophy of science, and in the 1903 edition of \textit{The Analysis of Sensations and the Relation of the Physical to the Psychical} explicitly referred readers to several of his articles in the journal. Ludwig Boltzmann also appeared in \textit{The Monist}; in 1901 it published two methodological essays by him on theoretical physics and the necessity of atomic theories. The journal likewise gave visibility to contemporary work in mathematical logic and foundations, publishing Felix Klein's ``The Present State of Mathematics'' in 1893, Oswald Veblen's ``Hilbert's Foundations of Geometry'' in 1903, and David Hilbert's ``On the Foundations of Logic and Arithmetic'' in 1905.

\subsection{Mathematical Density and the Monitoring of Science}

Beyond these major figures, \textit{The Monist} maintained a steady stream of mathematically oriented contributions on geometry, logic, and the foundations of arithmetic. These essays were not usually technical research papers in the later specialized sense. Even when written by mathematicians, physicists, or logicians, they were often conceptual, qualitative, and synthetic. They dealt with the foundations, meaning, methods, and philosophical interpretation of scientific knowledge in a form accessible to cultivated readers.

Another important feature of \textit{The Monist} was the substantial space devoted in each issue to reviews of recently published books. These review sections, often extending over many pages, discussed new works in mathematics, physics, chemistry, logic, and the philosophy of science. The authors reviewed before 1905 included J.~H. van 't Hoff, Wilhelm Ostwald, Ernst Mach, Hermann von Helmholtz, Ludwig Boltzmann, Pierre Duhem, A.~H. Bucherer, Felix Klein, Richard Dedekind, Jacques Hadamard, Louis Couturat, Edmund Husserl, and Karl Pearson. The review sections thus functioned as an intellectual observatory, tracking developments across the exact sciences and making European scholarly production visible to an international readership.

Published in Chicago but distributed through established European agents---most explicitly in London through Kegan Paul, Trench, Trübner and Co., Ltd.---the journal operated as a structured relay between continental scientific debates and the American academic environment. In this context, the January 1905 publication of Poincaré's Saint Louis lecture in \textit{The Monist} inserted the text into a transatlantic forum already attentive to Poincaré, Mach, Boltzmann, Hilbert, mathematical foundations, and the philosophical interpretation of modern science.

\section{Institutional Circulation of \textit{The Monist}, Vol.~15, No.~1 (1905)}

Evidence from library catalogues, accession registers, institutional stamps, and correspondence with library staff shows that the January 1905 issue of \textit{The Monist}, which contained the English translation of Poincaré's Saint Louis lecture, circulated rapidly within European academic institutions. The surviving record reveals a structured transatlantic distribution linking the United States to major European research libraries.

Institutional holdings of original 1905 copies of volume~15, distinct from the later 1966 Open Court reprint, are documented across a wide geographical area. The strongest evidence comes from dated institutional records. The copy of \textit{The Monist}, vol.~15, no.~1, in the Library of Congress / Smithsonian circuit bears a stamp dated 16 January 1905, only two weeks after publication. At Uppsala University Library, the January 1905 fascicle was formally registered on 3 February 1905, showing that it reached Sweden within only a few weeks of publication. At the University Library of Göttingen, the January 1905 issue was registered on 10 February 1905. According to archival information provided by the Staatsbibliothek zu Berlin, no.~1 of volume~15 was delivered to the Königliche Bibliothek Berlin on 24 February 1905 by the Harrassowitz book trade in Leipzig. The same source records regular deliveries for the remaining fascicles of the year: no.~2 on 9 May 1905, no.~3 on 19 July 1905, and no.~4 on 28 October 1905. These dates show that the Berlin copy formed part of an orderly current subscription supplied through an established German intermediary. Copies of the 1905 volume are also preserved at the Bayerische Staatsbibliothek in Munich, the Universitätsbibliothek Leipzig, the Universitätsbibliothek Erlangen--Nürnberg and the University Library of Strasbourg, then in the German Empire.

The Bodleian Library at Oxford preserves a continuous run of \textit{The Monist}; examination of volume~15 shows that the issues were received and processed at Oxford in the month of their publication. At the National Library of Finland in Helsinki, the 1905 issues were later bound together in a single volume bearing a 1906 binding stamp, a pattern consistent with regular subscription and receipt during or shortly after the year of publication. 

In France, the volume is held by the Bibliothèque nationale de France. At the Bibliothèque interuniversitaire de la Sorbonne, historical periodical registers show that \textit{The Monist} was being received by subscription in 1905. Additional original holdings are found at the Biblioteca Nazionale Centrale di Roma, the University of Cambridge and the Library of the Université catholique de Louvain. Surviving holdings in Stockholm, Göteborg, Lund, Copenhagen, and Helsinki further indicate that \textit{The Monist} belonged to the serial collections of major Nordic research libraries.

This evidence shows that the January 1905 issue was moving through university, national-library, and bookselling networks in the first months of 1905.

\subsection{Retail Availability of \textit{The Monist} in London and Paris}

The institutional circulation of \textit{The Monist} was complemented by commercial availability in major European book markets. The January 1905 issue itself printed both a London commercial relay and a British retail price: ``London: Kegan Paul, Trench, Trübner \& Company, Limited'' and ``In England and U.~P.~U., half a crown; Yearly, 9s.~6d.'' This indicates that the fascicle was explicitly prepared for sale on the British market, not merely distributed through institutional subscriptions.

Paris provides an equally important case. A contemporary French notice for \textit{The Monist} states: ``Bureau à Paris chez Brentano, 17 avenue de l'Opéra. Abonnement annuel : 12 francs. Prix d'un exemplaire, 3 francs.'' This gives direct evidence of a Parisian commercial point of sale, with subscription and single-issue prices in francs. Around 1905, therefore, the English translation of Poincaré's lecture was accessible not only through research libraries but also through retail channels in London and Paris. Retail access in other major European book centres was probably possible through the journal's international distribution network, although it is not directly attested by the documents presently available.

\section{\textit{The Bulletin des sciences mathématiques}:\\ Specialized Circulation}

By the beginning of the twentieth century, the \textit{Bulletin des sciences mathématiques} was a long-established and internationally recognized mathematical journal. Founded in 1870 under the title \textit{Bulletin des sciences mathématiques et astronomiques}, it was conceived not only as a venue for original work but also as an instrument of mathematical information, bibliographical monitoring, and scholarly communication. From 1885 onward it appeared under the shortened title \textit{Bulletin des sciences mathématiques}.

The publication of Poincaré's Saint Louis lecture in the December 1904 issue \cite{PoincareB} therefore opened a specialized channel of circulation, distinct from both \textit{La Revue des idées} and \textit{The Monist}. Whereas the former placed the lecture within the intellectual press, and the latter within a transatlantic forum for philosophical and scientific foundations, the \textit{Bulletin} inserted it into a professional mathematical network structured by research libraries, learned societies, and specialized subscriptions. This also helps explain why the \textit{Bulletin des sciences mathématiques} version later became the standard reference in much of the historiography.

Library catalogues show that runs including the December 1904 issue were preserved in a substantial number of European academic libraries. This evidence is important as an indication of the density of the institutional network through which the \textit{Bulletin} circulated. In Switzerland, holdings include ETH Zurich, the Bibliothèque de Genève, the University Library of Neuchâtel, the Université de Genève, and the Université de Fribourg. In Denmark, the journal is preserved both at AU Ny Munkegade in Aarhus, whose holdings cover the period from 1887 to 2008, and at KU Matematik in Copenhagen, where it is held from 1877 to 1997.

The same issue is traceable in major libraries of the German- and English-speaking academic worlds. In the German-speaking sphere, runs including 1904 are documented at Heidelberg, Nuremberg, the Bayerische Staatsbibliothek in Munich, the Technische Universität München, Berlin, Göttingen, and Strasbourg, then part of the German Empire. In the United Kingdom, holdings including 1904 appear in the catalogues of Oxford, Cambridge, the British Library, and University College London. In Italy, catalogue searches indicate holdings for 1904 at Bologna and at the Biblioteca del Dipartimento di Matematica ``G.~Castelnuovo'' of Sapienza University in Rome.

The ETH Zürich evidence is particularly valuable because precise arrival information of this kind is rarely preserved. The library was able to determine that the January 1904 issue entered its collections on 12 March 1904. Although this does not give the arrival date of the December 1904 fascicle itself, it shows that the \textit{Bulletin des sciences mathématiques} was reaching a major Swiss academic library as a current periodical on a timescale of roughly two months.

Direct evidence for the December 1904 fascicle containing Poincaré's lecture is available in Britain. The Bodleian Library copy at Oxford bears a stamp dated 13 January 1905, while the Cambridge copy bears an acquisition stamp dated 24 January 1905. These two traces show that the issue had reached two major British university libraries within only a few weeks of publication. They provide especially clear evidence that the \textit{Bulletin} version of Poincaré's lecture was already present in major academic collections before the end of January 1905.

The geographically dispersed holdings of the \textit{Bulletin des sciences mathématiques}, combined with the dated evidence from Oxford and Cambridge and the ETH indication of regular serial receipt, show that Poincaré's lecture entered a dense and specialized European infrastructure of mathematical communication. Its December 1904 publication was therefore not only a republication in a learned journal, but a movement into a professional transnational circuit extending from France to Switzerland, Germany, Britain, Denmark, and Italy.

\section{\textit{La valeur de la science}: Publication, Circulation, and Early Reception}
Before the present investigation, the most readily accessible online bibliographical resources indicated only the year 1905 for the publication of \textit{La valeur de la science} \cite{Poincare1905}. The evidence assembled here makes it possible to refine this chronology considerably. It indicates that the volume was published in the spring of 1905 and entered circulation very rapidly.
An extract from the book appeared in \textit{La Nouvelle Revue} in Paris on 15 April 1905, under the title ``La valeur de la science.'' A note accompanying the text stated that these pages were taken from the volume \textit{La valeur de la science}, which was to appear shortly with the publisher Ernest Flammarion. The extract was not a brief promotional notice but a substantial portion of the book's Introduction. It already presented the programmatic themes of the volume, including the value of scientific truth, the relation between \textit{La valeur de la science} and \textit{La science et l'hypoth\`ese}, the announcement of a discussion of time and space as the ``frames'' within which nature appears to us, and an explicit reference to Poincar\'e's Saint Louis lecture of 1904. Its separate publication is therefore especially significant. It shows that the intellectual programme of the book, together with its explicit connection to the Saint Louis lecture, was already in print and publicly visible before the volume itself was widely advertised.

A particularly early Belgian marker is provided by \textit{Le Petit Bleu du Matin} of Brussels, which announced the publication of \textit{La valeur de la science} on 30 April 1905. In Paris, the \textit{Journal des débats} listed the book under \textit{Publications récentes} on 13 May 1905. These elements show that the volume was publicly visible by the end of April and commercially available in Paris by mid-May at the latest.

A second Belgian witness, published only one day after the notice in \textit{Le Petit Bleu du Matin}, is even more revealing. On 1 May 1905, \textit{Le Soir} of Brussels published a first-page article signed ``PICCOLO,'' very probably Auguste Cauvin d'Arsac, a central figure at \textit{Le Soir} and editor-in-chief of the newspaper. This was not merely a bookseller's advertisement or a bibliographical notice. Poincaré's book was used in an autonomous reflection on the value of scientific theories, their apparent fragility, and their practical effectiveness. The article cited a precise passage from \textit{La valeur de la science}, concerning the comparison between science and the rules of a game such as tric-trac. It therefore documents not only commercial availability, but an early act of reading, quotation, and intellectual appropriation in Brussels by 1 May 1905.

A further indication of very early international awareness is provided by the letter sent by Gösta Mittag-Leffler to Henri Poincaré on 30 April 1905 \cite{MittagLefflerPoincare1905}, in which Mittag-Leffler refers to \textit{La valeur de la science} and asks Poincaré to offer a copy to the King of Sweden. The point is all the more significant because the book was not mentioned in Poincaré's preceding letter to Mittag-Leffler, so that its appearance in the correspondence on 30 April marks a very narrow and early window of recognition. This document shows that the book had already become known in an international scholarly milieu by the end of April. Taken together with the Belgian evidence, it confirms that the circulation of the volume began almost immediately and was not confined to the Parisian book market.

Material evidence from Göttingen reinforces the same conclusion. The University Library inventoried \textit{La valeur de la science} on 12 May 1905, recording that it had been supplied by A.~Asher and Company, Buchhandlung und Verlag, Berlin. This is a particularly significant indication, since it shows that the book had already entered ordinary international bookselling channels and had reached one of Germany's leading academic centres by mid-May 1905.

Swiss evidence points in the same direction. The May 1905 issue of the \textit{Revue historique vaudoise}, vol.~13, carried an advertisement by the Lausanne bookseller Payot et Compagnie announcing Poincaré's \textit{La valeur de la science} as ``vient de paraître.'' The \textit{Bulletin technique de la Suisse romande}, vol.~31, no.~10, of 25 May 1905, also carried an advertisement by the Lausanne bookseller F.~Rouge and Cie. These notices show that the book had entered the Swiss commercial book trade before the end of May 1905.

\subsection{A Geneva Scientific Column: \'{E}mile Yung on \textit{La valeur de la science}}

The Swiss reception of \textit{La valeur de la science} was not limited to advertisements and bibliographical notices. On 26 June 1905, the \textit{Journal de Genève} published a substantial first-page article by \'{E}mile Yung in its literary and scientific issue, under the rubric ``Science et Nature.'' Yung was not a theoretical physicist or mathematician, but a Geneva zoologist, professor, and recognized scientific intellectual. His article is therefore an important witness to the rapid diffusion of Poincar\'e's book beyond the narrow circle of specialists in mathematical physics.

The position of the article within Yung's column makes the evidence still more significant. The rubric ``Science et Nature'' appears, on the basis of the articles so far identified, to have begun only a few weeks earlier, on 22 May 1905, with Yung's article ``La diss\'emination des araign\'ees,'' a subject directly connected with his own field as a zoologist and naturalist. No earlier instance of the rubric has been found in the issues of the \textit{Journal de Gen\`eve}. The article on \textit{La valeur de la science} was therefore the second article identified in a newly opened scientific column, not one item in a long-established routine of scientific reviewing.

This context is important because the later articles of 1905 identified in the same rubric largely correspond to Yung's ordinary scientific interests. All belong broadly to zoology, biology, physiology, or naturalist observation. Against this background, the article on Poincar\'e stands out as an exceptional intervention. It took Yung outside the usual domain of the column and into questions of scientific philosophy, mathematical physics, and the foundations of knowledge. The prominence given to \textit{La valeur de la science} at such an early moment in the history of the rubric suggests a deliberate editorial choice rather than a merely routine notice prompted by the arrival of a publisher's circular or review copy.

The date of publication, 26 June 1905, is itself only the visible endpoint of a prior process of access, reading, composition, and editorial preparation. Yung would already have had to obtain and read the book, judge it worthy of public discussion in a newly established scientific column, compose his article, and submit it to the newspaper before that date. The article therefore documents the presence and active reception of \textit{La valeur de la science} in Geneva during the preceding weeks.

The content of Yung's article is especially revealing. He presents the book in relation to the contemporary crisis of the fundamental principles of physics, including the apparent fragility of principles such as the conservation of energy and the conservation of mass. He does not treat Poincar\'e's work as an abstract philosophical essay detached from current scientific concerns, but as an intervention in a living debate about the foundations of science. Yung also identifies the notions of time and space as among the central ``frames'' within which nature appears to us. Poincar\'e's reflections on time and space were therefore visible enough to be extracted and presented in a general scientific column in a major Swiss newspaper, and visible at precisely the moment when the book had only recently entered public circulation.

Having established this important Geneva witness, we can now return to the broader evidence for the circulation of \textit{La valeur de la science} before the end of June 1905.

These elements considerably strengthen the chronology of early circulation. By the end of April 1905, \textit{La valeur de la science} was already publicly announced in Brussels and known in an international scholarly milieu. By 1 May it was cited in the Brussels press. By 12 May it had reached Göttingen through international bookselling channels. By May it was advertised in French-speaking Switzerland. By 26 June, after an earlier process of access, reading, composition, and editorial transmission, it was the subject of a substantial interpretive article in Geneva. The book was therefore not only published in the spring of 1905. It was rapidly distributed, advertised, acquired, cited, read, and discussed in several European contexts.

A further indication that French books by Poincaré circulated through ordinary bookshop channels in Einstein's Bernese milieu is provided by Carl Seelig's account of the Olympia Academy. Seelig, who was then preparing a biography of Einstein, reports that Conrad Habicht brought Henri Poincaré's \textit{La science et l'hypothèse} \cite{Poincare1902} from a bookshop as a much-discussed new publication. This detail deserves attention. It presents Poincaré's book as a commercially visible work that could be bought in a local bookshop and introduced into the reading circle of Einstein, Habicht, and Solovine.

The documentary background of Seelig's statement strengthens its value. In February and March 1952 Seelig actively sought information from Habicht about his relations with Einstein, after Besso had identified Habicht as one of Einstein's close companions from the Swiss years. Seelig then visited Habicht, consulted the letters, cards, and envelopes that Habicht placed at his disposal, and asked him to check passages in the correspondence that he could not decipher with certainty \cite{SeeligHabicht1952}. Seelig's later statement about Habicht bringing Poincaré's book from a bookshop should therefore be read against the background of direct contact with Habicht and of a material consultation of Einstein's correspondence with him, not merely as a loose literary reconstruction.

The chronology also suggests that the work read in the Olympia Academy was the French edition of \textit{La science et l'hypothèse}, published by Flammarion in 1902, rather than the later German translation. Einstein's postcards to Habicht show that by late 1903 and early 1904 Habicht was no longer continuously present in Bern \cite{EinsteinH}. If Seelig's episode belongs, as seems most natural, to the early phase of the Olympia Academy, then the book brought by Habicht was most probably the French Flammarion edition. This is significant for the present argument. It shows that French philosophical-scientific works by Poincaré were not outside the commercial horizon of Einstein's Bernese circle. Consequently, when \textit{La valeur de la science} appeared in 1905 from the same publisher and was rapidly advertised and distributed, it is entirely plausible that it too could have been obtained through the same ordinary channels, whether by Solovine or by another member of the Bernese intellectual circle around Einstein.

\section{Redating the Habicht Letter: May or June 1905?}

The dating of Einstein's well-known letter to Conrad Habicht is more important than it may first appear. The letter is not merely a private note. It is a chronological hinge between the circulation of Poincaré's \textit{La valeur de la science}, the final phase of Einstein's work on the electrodynamics of moving bodies, and the surviving evidence concerning his Bernese intellectual environment. In announcing four works to Habicht, Einstein describes the fourth as a paper on the electrodynamics of moving bodies, still only ``in draft''. The relativity paper was therefore not yet completed, but it was already sufficiently formed to be presented as a distinct work involving a modification of the theory of space and time.

The usual editorial dating places the letter in Bern on Thursday 18 or 25 May 1905. This dating rests on the fact that the letter is written on a Thursday and that Einstein sends greetings from his wife and from Hans Albert, who was now one year old. Since Hans Albert was born on 14 May 1904, the letter must have been written after 14 May 1905. The two Thursdays immediately following this date were 18 and 25 May. Yet these are not the only internal possibilities. Thursday 1 June 1905 also satisfies the evidence given by the letter itself.

A further clue favours a date close to the formal appearance of Einstein's first paper, the light-quantum article. In the Habicht letter Einstein writes that he may soon send Habicht this first paper, since he will soon receive the complimentary reprints. The article on radiation and the energy properties of light appeared on 9 June 1905. Since complimentary offprints could reach authors shortly before or around the formal appearance of an issue, Einstein's remark fits particularly well with a letter written close to 9 June, and therefore with Thursday 1 June rather than with 18 May.

The biographical context also makes an immediate 18 May dating very strained. Einstein's paper on Brownian motion was received by the \textit{Annalen der Physik} on 11 May. If the transmission followed the same pattern as the light-quantum paper, which was received by the journal one day after dispatch, the Brownian-motion manuscript must in all likelihood have been sent on 10 May. At almost the same time, the Einsteins were also preparing a domestic move, which must have involved at least some search, arrangement, and practical organization. The lease for the new family apartment was registered on 13 May \cite{HentschelGrasshoff2005}; Hans Albert turned one on 14 May; and, according to Flückiger \cite{Fluckiger1974}, the family moved on 16 May from Kramgasse 49, in the old town, to Besenscheuerweg 28, now Tscharnerstrasse 28. The new address was in a more peripheral residential area to the west or south-west of the historic centre, in the direction of the Mattenhof--Monbijou district. This was not a simple domestic event for a young family with a one-year-old child, and Einstein was still working full time at the Patent Office. A letter written on 18 May would therefore fall only two days after the move and only one week after the Brownian-motion paper had been received by the \textit{Annalen}. The sequence would therefore be exceptionally compressed.

This compression is important because the Habicht letter does not merely date an unfinished manuscript. It also helps to locate the final phase of Einstein's work on electrodynamics within his immediate Bernese environment. If the letter belongs not to 18 May but to 25 May or 1 June, then the relevant period falls after Einstein's move from Kramgasse 49 to Besenscheuerweg 28. At that point the question of Einstein's practical proximity to Michele Besso becomes directly relevant. Besso was not only a colleague at the Patent Office; he was also the person whom Einstein later associated with the decisive discussion that helped him overcome the difficulty in the relativity problem. The reconstruction of Besso's address is therefore part of the same chronological question.

The relevant local setting can be reconstructed only within certain limits from contemporary Bernese administrative records. The Bern address book for 1904/05 lists Michele Angelo Besso, engineer at the Federal Office for Intellectual Property, at Zeughausgasse 41, about 300 metres from the Patent Office \cite{SAB_D5_24}. The residents' register of the Stadtarchiv Bern records Besso at Schwarzenburgstrasse 15 \cite{SAB_1125_19_2}, and the following Bern address book, for 1905/06, also lists him at Schwarzenburgstrasse 15 \cite{SAB_D5_25}. These documents establish a change of address from Zeughausgasse 41 to Schwarzenburgstrasse 15 in 1905, but they do not fix the exact date of Besso's physical move. It would therefore be unsafe to infer, solely from these records, that Besso had already moved to Schwarzenburgstrasse 15 before Einstein's own move from Kramgasse 49 to Besenscheuerweg 28 on 16 May 1905.

Einstein's move nevertheless altered the practical geography of his daily life in a way that is relevant to the chronology. From Kramgasse 49, a regular walk back from the Patent Office together with Besso would have been less natural, since Einstein and Besso lived in different directions from the office. After Einstein settled at Tscharnerstrasse 28, however, the walk from his new address to the Patent Office took approximately 20--25 minutes. If Besso had already moved to Schwarzenburgstrasse 15, Einstein then lived only about 250 metres from Besso's new address. In that case, informal encounters, shared walks, and evening exchanges would have become materially easier than before. If Besso was still living at Zeughausgasse 41 for some time after 16 May, the idea of a shared homeward route would be topographically less compelling. The point is therefore that the second half of May and the beginning of June offer a more plausible practical setting for the remembered pattern of conversations than the already compressed days before Einstein's move.

This point is supported by Einstein's own later references to his conversations with Besso on the way home. The earliest explicit surviving reference occurs in Einstein's letter to Maurice Solovine of 27 April 1906 \cite{CPAE5Eng}, only a few months after the decisive period. In that letter Einstein wrote that he had moved again, this time back to the Kirchenfeld district, to Aegertenstrasse 53, about two kilometres from the area in which Besso was then living. Einstein immediately added that even the conversations with Besso on the way home had now come to an end. The remark shows that these homeward conversations had already become a familiar and memorable feature of Einstein's Bernese life, and that Einstein himself associated their interruption with a change in the practical geography of his daily movements.
The same motif reappears much later. In 1952, when Besso asked Einstein what he should tell Seelig about the Bern years, Einstein again mentioned their conversations on the way home \cite{Speziali1972}; and after Besso's death, in his letter of 21 March 1955 to Besso's family \cite{Speziali1972}, Einstein recalled that these conversations on the way home had an incomparable charm, as if human contingencies did not exist at all. These repeated references, from 1906 to 1955, give the homeward conversations a special evidential weight. They were a remembered practice embedded in the daily geography of Bern.

This proximity is especially relevant in view of Einstein's later recollection of the decisive discussion with Besso. Einstein stated that, after having struggled in vain with the problem, he went to visit Besso in order to discuss it with him. The next day, according to this recollection, Einstein returned to Besso and was able to say that he had completely solved the problem. This retrospective account gives particular significance to the material conditions under which such an exchange could have taken place. If the decisive phase of Einstein's work on electrodynamics occurred in the second half of May or around the beginning of June, the new practical geography after Einstein's move becomes directly relevant to the circumstances behind Einstein's acknowledgment, at the end of the published relativity paper, that he was indebted to Besso for several valuable suggestions.

These elements show that the traditional alternatives of 18 and 25 May for the date of the letter should not be treated as exhaustive. A date of 1 June 1905 satisfies the internal evidence of the letter, fits particularly well with Einstein's remark about the expected reprints of the light-quantum paper, avoids the extreme compression produced by an 18 May dating, and places the letter in a biographical and topographical setting in which Einstein's intensified exchanges with Besso are materially easier to understand.

The chronological issue intersects directly with the circulation of Poincaré's \textit{La valeur de la science}. As shown above, the book was already publicly visible in Belgium by the end of April 1905, cited in the Brussels press on 1 May, and known to Mittag-Leffler by 30 April. It had reached Göttingen by 12 May through the Berlin bookseller A.~Asher and Company, and it was advertised in French-speaking Switzerland before the end of May. These traces make its availability through Swiss or German book channels materially plausible in the weeks preceding Einstein's submission of the relativity paper.

The Habicht letter is therefore a chronological hinge. If it is dated 18 May, the interval between the earliest documented circulation of \textit{La valeur de la science} and Einstein's draft of the electrodynamics paper becomes extremely narrow. If it is dated 25 May, the interval becomes more plausible. If 1 June is admitted as an internal possibility, the chronology becomes considerably less compressed. The point is that the availability of Poincaré's book precedes, or at least overlaps with, the earliest attested stage of Einstein's electrodynamics manuscript.

In such a compressed chronology, a difference of one or two weeks is not negligible. It affects how one evaluates the relation between the circulation of Poincaré's book and the formation of the 1905 relativity paper. It does not prove that Einstein read Poincaré's book before completing his paper, but it changes the chronological conditions under which that question has to be assessed.

\section{The Olympia Academy, Solovine's Testimony, and the Narrowing of Later Memory}

The dating of the Habicht letter also matters because it determines how close Einstein's four-paper announcement stands to the surviving evidence concerning the Olympia Academy. Two earlier letters from Einstein to Habicht, both dated by the editors to Monday 6 March 1905, are important for this immediate context. In the first, Einstein urged Habicht not to forget Bern and the Academy and asked him to come and see them. In the second, he jokingly beseeched, warned, and ordered Habicht to attend some of the sessions of their ``estimable academy'' and thereby to increase its current membership by fifty percent \cite{CPAE5Eng}. These two short notes show that Habicht was no longer regularly present in Bern by March 1905. At the same time, they show that the Academy had not simply disappeared. Einstein still referred to its sessions as continuing, and the joke about increasing the membership by fifty percent suggests that the active group had been reduced to two persons, almost certainly Einstein and Solovine.

This gives particular weight to Solovine's later recollections. They were not memories of a circle that had necessarily ceased to exist before the decisive months of 1905, but of an intellectual setting that appears to have continued, in reduced form, on the eve of Einstein's four-paper announcement. 

Solovine's correspondence with Seelig is equally important for the intellectual context of the Olympia Academy. In the letter of 14 April 1952, Solovine described the group's evening discussions and named works by Mach, Mill, Hume, Dedekind, and both \textit{La science et l'hypothèse} and \textit{La valeur de la science} by Poincaré \cite{SolovineSeelig1952}. This retrospective list is not chronological and cannot by itself date a reading before June 1905. But it shows that, in Solovine's memory, the two books of Poincaré belonged naturally to the shared intellectual repertoire of the group. The point is all the more significant because Solovine's three other known letters to Seelig preserved at the ETH Archives, the last of them dating from 1955, do not correct or retract this list of readings \cite{SolovineSeelig1952b,SolovineSeelig1952c,SolovineSeelig1955}.

The manuscript testimony received by Seelig was therefore richer than the later printed accounts. Despite Solovine's explicit mention of \textit{La valeur de la science} in April 1952, Seelig did not include this title in his published accounts of the Olympia Academy readings in 1952, 1954, or in the English version of 1956 \cite{Seelig1952,Seelig1954,Seelig1956}. The omission is noteworthy because Seelig was not working at a great distance from the surviving private documentation. He had visited Conrad Habicht at his home, and Habicht had lent him the letters he had received from Einstein. Seelig thus had direct access both to Solovine's retrospective testimony and to the Habicht correspondence. Yet the printed biographical narrative that emerged from this material gave a narrower picture of the group's readings than Solovine's private letter had supplied.

The same narrowing can also be observed on Solovine's side. In the introduction to his 1956 edition of Einstein's letters to him, Solovine no longer included \textit{La valeur de la science} among the works read by the Olympia Academy \cite{Solovine1956}. The omission reflects a broader passage from private recollection to printed memory, in which the documentary record became more selective. This does not prove when \textit{La valeur de la science} was read, and it does not establish that it was read before June 1905. But it shows that the later published tradition simplified a richer manuscript testimony in which Poincaré's 1905 book still formed part of the remembered intellectual repertoire of the Olympia Academy.

The significance of this point is chronological as well as historiographical. By itself, Solovine's 1952 letter cannot establish that the Olympia Academy read \textit{La valeur de la science} before Einstein completed his relativity paper. But once the early circulation of the book has been reconstructed, the testimony acquires a different status. Solovine's memory that it belonged to the Academy's readings is therefore not incompatible with the documented circulation of the book in the relevant period. If the Habicht letter is dated to 25 May or 1 June rather than 18 May, the overlap between the book's early European circulation, the reduced but continuing Academy, and Einstein's still unfinished electrodynamics paper becomes still more concrete.

The manuscript record preserves a broader memory than the published biographies transmitted. In combination with the documented circulation of Poincaré's book and the reopened chronology of the Habicht letter, it shows that the question of possible access and reading must be posed within a denser and less isolated Bernese documentary environment than the later simplified narrative suggests.

\section{Joseph Sauter's Testimony on Einstein, Lorentz, and the Patent Office Milieu}

A further testimony deserves attention in this context: Joseph Sauter's account of his relations with Einstein at the Swiss Federal Office of Intellectual Property. Sauter belonged to Einstein's immediate professional environment at the Patent Office and was scientifically more qualified than most of Einstein's colleagues there. A former student at the Zurich Polytechnic, assistant of Heinrich Friedrich Weber, and doctor in physics, he had published in leading German-language journals: in 1897 on the magnetization of a ring in the \textit{Annalen der Physik und Chemie}, and in 1901 on the interpretation of Maxwell's equations in resting isotropic media in the \textit{Annalen der Physik}. In Einstein's Bern surroundings, only Michele Besso was comparably important as a technical interlocutor. Sauter was therefore not merely a friendly colleague to whom Einstein could give a simplified explanation, but a trained physicist and engineer familiar with Maxwellian electrodynamics.

Sauter's testimony originated in his participation in a radio broadcast in Bern in 1955 and was later prepared from the recording of that broadcast. It was published in 1966 in the internal journal of the Swiss Federal Office of Intellectual Property, \textit{Echo du Bureau}, under the title ``Comment j'ai appris à connaître Einstein'' \cite{Sauter1966}. The copy used here was kindly supplied by Alexandra Graber, Specialist in Information and Documentation at the Swiss Federal Institute of Intellectual Property, Bern. Although retrospective, the testimony has an unusually significant provenance: it comes from the institutional milieu of the Patent Office itself and preserves the recollection of one of the few scientifically competent colleagues with whom Einstein discussed his work in Bern. The same text was later reproduced by Max Flückiger in his documentary book on Einstein in Bern \cite{Fluckiger1974}. Parts of the testimony were also reproduced by Alberto Martínez \cite{Martinez2009}. Sauter's account of Einstein's explanation of clock synchronization by light signals, illustrated by the towers of Bern and Muri, has been mentioned in several publications. That episode, however, is not the focus of the present discussion. The more significant point for the present argument is Sauter's recollection of Einstein's reaction to the agreement between his results and Lorentz's formulae. This part of the testimony appears not to have been analyzed in the historical literature and has only been mentioned in passing by Martínez.

Sauter's status as a witness was also recognized in 1955, when he was invited to the Bern congress organized for the fiftieth anniversary of special relativity. This shows that he was regarded in the commemorative context itself as a legitimate surviving witness of Einstein's Bern years. His testimony belongs to the very small group of reminiscences coming from persons who had actually known Einstein in the Patent Office environment and who were still treated, half a century later, as relevant witnesses of that milieu \cite{MercierKervaire1956}. Its lateness must be kept in mind. Yet this very limitation also gives the document a particular importance: within the presently known body of testimonies from persons who were close to Einstein in Bern in 1905, Sauter's account appears to be the only one that reports a direct recollection of Einstein discussing the relativity paper.

Sauter first situates Einstein within the culture of the Patent Office. He recalls Einstein's appointment in 1902 as technical expert and emphasizes that Einstein was known there as a physicist familiar with Maxwell's theory. Sauter also presents himself as someone who had already been interested in Maxwellian theory and who had published a mechanical model leading to differential equations of the same form as Maxwell's equations. This is important because it shows that the discussions between Sauter and Einstein were not casual conversations between a physicist and a layman. When Sauter reports Einstein's explanations, he is reporting them from the standpoint of a colleague who possessed relevant technical competence.

This point is reinforced by Sauter's account of their conversations outside office hours. He states that his best discussions with Einstein took place when they were returning home and that Einstein made him read and understand his most recent scientific publications. Sauter adds that Einstein began with the thermodynamic papers, that is, with the statistical-mechanical works preceding the relativity memoir. He even reports that Einstein acknowledged before him an error in one of these works and refused to be consoled. This passage places Sauter inside a real scientific exchange with Einstein before the discussion of relativity. He was not merely recalling a single exceptional conversation about the famous paper of 1905, but a broader pattern of discussions in which Einstein explained to him his recent research.

The significance of this point is strengthened by Einstein's own later recollection. In a letter to Michele Besso of 13 July 1952, written in response to a question from Besso, Einstein confirmed that he had had extensive discussions with Sauter about his early statistical-mechanical works \cite{Speziali1972}. This independent confirmation is methodologically important. It means that, on at least one central feature of Sauter's reminiscence---the fact that Einstein discussed his early theoretical papers with him in some detail---Sauter's account is corroborated by Einstein himself.

Only after this preliminary discussion of the thermodynamic and statistical papers does Sauter turn to the 1905 relativity memoir. He writes that, after the thermodynamic papers, there came in June 1905 the celebrated memoir on special relativity. He describes Einstein as beginning from serious criticisms of the foundations of classical electromagnetism, especially the failure of the Michelson experiment. According to Sauter, classical electromagnetism would have led one to expect a displacement in Michelson's experiment, whereas no such displacement was observed. Sauter therefore remembered Einstein's exposition of relativity as directly connected with the crisis of electromagnetic theory and with the negative result of Michelson and Morley.

A passage in Sauter's testimony is especially significant. Sauter reports that Einstein experienced an ``indescribable joy'' when, after completing the memoir, he noticed that his results agreed with the formulae found by Lorentz. He even quotes Einstein as saying that he could not express the intensity of his joy. This is an important piece of evidence because it shows that Lorentz's formulae were present in Einstein's horizon at the decisive moment of completion or immediate verification of the paper. The passage does not decide how Einstein knew Lorentz's results. It does not prove that he had read Lorentz's 1904 memoir in full; nor does it exclude the possibility that he knew the formulae through summaries, discussions, reviews, or other indirect channels. But it makes difficult any simple narrative of complete documentary ignorance of Lorentz's relevant results at the moment when Einstein finished the work.

This conclusion is reinforced by the contemporary visibility of Lorentz's 1904 memoir. Poincaré discussed Lorentz's theory in his Saint Louis lecture of September 1904. In July 1904, only a few months after Lorentz's memoir, Wien referred to Lorentz's new theory and wrote down the relevant transformation formulae explicitly in a short article in the \textit{Physikalische Zeitschrift} \cite{Wien1904}. These included both the transformation of the space-time variables and the corresponding transformation of the electromagnetic field components. In the same year, Bucherer's \textit{Mathematische Einführung in die Elektronentheorie}, whose preface is dated July 1904, also incorporated Lorentz's recent memoir, discussed the contraction hypothesis, and presented Lorentz's transformation, together with the associated field transformations, within a German textbook devoted to electron theory \cite{Bucherer1904}. These examples show that Lorentz's 1904 results were not accessible only through the original memoir. Before Einstein submitted his relativity paper, central elements of Lorentz's theory had already been restated in concise German-language publications, including publications available in Bern. In February 1905, Richard Gans published a review of Lorentz's memoir in the \textit{Beiblätter zu den Annalen der Physik} \cite{Gans1905}. Langevin presented Lorentz's work in early 1905 \cite{Langevin1905}, and related references soon appeared in writings by Abraham \cite{Abraham1904}, Cohn \cite{Cohn1904}, and Sommerfeld \cite{Sommerfeld1904a, Sommerfeld1904b, Sommerfeld1905} on electron theory and the electrodynamics of moving bodies. Sauter's testimony should therefore be read against a scientific background in which Lorentz's 1904 results were already known, discussed, summarized, and cited. It does not settle the precise channel through which Einstein may have become familiar with them, but it makes a narrative of documentary isolation increasingly difficult to sustain.

Sauter immediately adds an important nuance: Einstein's advantage over Lorentz was that he did not need the Fitzgerald contraction hypothesis in the same explanatory role. Sauter's memory preserves precisely the distinction that matters historically. Einstein's results coincided formally with Lorentz's formulae, and this agreement gave Einstein intense satisfaction; yet Einstein's conceptual route was different, because the contraction no longer functioned as an ad hoc dynamical hypothesis introduced to save the negative result of Michelson and Morley.

Sauter's testimony is a retrospective source and must be treated as such. It was written fifty years after the events and cannot be used to reconstruct every detail of the conversations with certainty. But it is not an isolated anecdote without external support. The part of Sauter's account concerning discussions of Einstein's early statistical-mechanical works is independently supported by Einstein's own 1952 recollection to Besso. Moreover, Sauter was not a marginal or technically unqualified witness: he belonged to the Patent Office milieu, had discussed Einstein's recent theoretical work with him, and had the scientific training needed to understand the Maxwell-Lorentz background. The testimony is therefore valuable because it preserves a rare recollection from Einstein's immediate Bern environment. In particular, its neglected Lorentz passage shows that, at the moment of completion or immediate verification of the 1905 memoir, Einstein's relation to Lorentz's formulae was not one of simple absence or irrelevance, but one of recognized and emotionally powerful concordance.

\section{Major Contemporary Physics Publications Available in Bern}

By 1905, the holdings of the University Library of Bern covered several of the most important channels through which contemporary physics was published, reviewed, abstracted, and circulated. The catalogue records of the Universität und PH Bern, as listed in Swisscovery, show that the library held the main German, French, Dutch, British, and international periodicals relevant to the physics of the period. These included the \emph{Annalen der Physik}, with holdings covering 1824--1993; the \emph{Beiblätter zu den Annalen der Physik}, with holdings covering 1900--1919; the \emph{Physikalische Zeitschrift}, with holdings covering 1899--1944; the \emph{Verhandlungen der Deutschen Physikalischen Gesellschaft}, with holdings covering 1899--1919; the \textit{Nachrichten von der Königlichen Gesellschaft der Wissenschaften zu Göttingen, Mathematisch-physikalische Klasse}, with holdings covering 1895--1933; the \emph{Comptes rendus hebdomadaires des séances de l'Académie des sciences}, with holdings covering 1846--1950; the \emph{Proceedings of the Section of Sciences} of the Royal Academy of Amsterdam, listed in Swisscovery as \emph{Actes de la section des sciences}, with holdings covering 1898--1937; \emph{Nature}, with holdings covering 1869--2016; and \emph{The London, Edinburgh, and Dublin Philosophical Magazine and Journal of Science}, with holdings covering 1858--1903 and 1905--1925.

The significance of these holdings is not merely institutional. They correspond to the publication landscape in which the problems central to Einstein's work between 1902 and 1905 were being discussed. The \emph{Annalen der Physik} was the principal German journal of physics and the journal in which Einstein himself published his 1905 papers. The associated review periodical, the \emph{Beiblätter zu den Annalen der Physik}, held in Bern for the years 1900--1919, constituted an important bibliographical instrument, allowing German-reading physicists to follow recent publications beyond the articles they happened to read directly. The \emph{Physikalische Zeitschrift}, the \emph{Nachrichten von der Königlichen Gesellschaft der Wissenschaften zu Göttingen, Mathematisch-physikalische Klasse} and the \emph{Verhandlungen der Deutschen Physikalischen Gesellschaft} further supplied access to current discussions in German physics.

The French, Dutch, British, and international channels were also represented. The Bern holdings of the \emph{Comptes rendus hebdomadaires des séances de l'Académie des sciences} covered the crucial year 1905. This journal was one of the principal European channels for rapid scientific communication. The library also held the \emph{Proceedings of the Section of Sciences} of the Amsterdam Academy for the period concerned. This is especially important because Lorentz's 1904 memoir, ``Electromagnetic phenomena in a system moving with any velocity smaller than that of light,'' appeared in that series, in volume~6. The Bern catalogue therefore shows that the series containing Lorentz's paper belonged to the local documentary environment available for the period in question.

The presence of \emph{Nature} and the \emph{Philosophical Magazine} is also significant. \emph{Nature} was not only a general scientific weekly but also a rapid international channel for notices, reports of meetings, reviews, and summaries of work published elsewhere. Its presence in Bern meant that British and international scientific news, including reports of congresses and society meetings, formed part of the periodical environment available locally. The \emph{Philosophical Magazine}, one of the central British journals for physics, was likewise represented in the Bern holdings. Although the catalogue record shows a gap for 1904, the journal was held for the long period 1858--1903 and again from 1905 onward. The availability of these British periodicals is relevant because they formed part of the broader international circulation of physical theory, molecular physics, electrodynamics, and related experimental and mathematical work at the turn of the century.

Nor was the Bern collection limited to periodicals. The library also held, for instance, the large collective volume \emph{Festschrift Ludwig Boltzmann: gewidmet zum sechzigsten Geburtstage, 20. Februar 1904}, published by J.~A.~Barth in Leipzig in 1904. This nearly thousand-page volume brought together 117 contributions by leading figures in mathematics, physics, physical chemistry, and related fields. In this respect, although on a different scale and in a different format, the volume performed a function somewhat analogous to that of the large international congresses around 1900: it gathered a dispersed scholarly community into a single printed object and made visible, through publication, the range and density of contemporary scientific work. It included, among other things, Planck's paper on the mechanical meaning of temperature and entropy, in which Gibbs's statistical definitions of entropy were discussed in detail, and Smoluchowski's paper on irregularities in the distribution of gas molecules and their influence on entropy and the equation of state, one of the most substantial treatments of molecular fluctuations of the period. The significance of the volume is further reinforced by Einstein's own reviewing activity. In the first half of March 1905, Einstein reviewed in the \emph{Beiblätter zu den Annalen der Physik} several contributions from the same \emph{Festschrift}, including papers by George Hartley Bryan, Nikolay Nikolayevich Schiller, and Jacobus Henricus van 't Hoff. The volume therefore illustrates not only the breadth of Bern's contemporary scientific holdings, but also the kind of recent international literature with which Einstein could come into contact through his bibliographical work.

These holdings make clear that the relevant literature was not inaccessible in Bern. For this reason, the question of Einstein's knowledge of the contemporary literature should not begin from an assumption of documentary isolation. The more historically appropriate question is which of these available channels he may have used in the years immediately preceding June 1905.

The Hochschulbibliothek was listed at Herrengasse 38, beside the old university. Its reading rooms were open from 10 a.m. to noon and from 2 p.m. to 7 p.m., until 8 p.m. in winter \cite{ub-bern-oeffnungszeiten}. Recent periodicals and reference works were not only part of the library's holdings; they were available for on-site consultation in reading rooms.

\subsection{Einstein's Reading Practices and the Search for Relevant Literature in Bern between 1902 and 1905}

Einstein's correspondence from the years preceding 1905 does not support the image of a young physicist indifferent to current scientific literature. It shows instead a researcher who actively sought books, papers, references, tables, and offprints through friends, colleagues, libraries, and personal networks. When a topic mattered to his work, Einstein looked for relevant literature and used the documentary channels available to him.

An early example appears in his correspondence with Mileva Marić. In a letter of 28 December 1901, some weeks before his arrival in Bern, Einstein wrote that he intended to ask his former classmate Jakob Ehrat to send him the latest works of Drude and Lorentz \cite{CPAE1Eng}. Ehrat had studied at the Zurich Polytechnic in the same cohort as Einstein and, after graduating in 1900, remained there as an assistant. The remark is significant because Drude and Lorentz were among the central authors for a young physicist interested in electron theory, optics, and the electrodynamics of moving bodies. It shows that Einstein was ready to use personal academic contacts in order to obtain recent scientific literature.

This was not an isolated gesture. In a letter to Ehrat written in late March 1903, Einstein again referred to scientific material obtained through his former Zurich contact. He complained, however, that the tables Ehrat had sent him were of no use, since they dated from 1883 \cite{CPAE5Eng}. The remark is revealing. Einstein was not merely asking for documents; he was asking for documents that were sufficiently recent to be scientifically useful. The same letter also shows that Ehrat remained part of Einstein's scientific network. Einstein informed him that his latest paper on the foundations of thermodynamics would soon appear, described it as deriving the concepts of temperature and entropy, as well as the second law, from the energy principle and atomistic theory, and invited Ehrat to read and judge it \cite{CPAE5Eng}. Ehrat thus appears not only as a supplier of material from the Polytechnic, but also as someone to whom Einstein could still present ongoing research.

The same pattern appears, in a more substantial form, in Einstein's exchanges with Michele Besso, who was then in Trieste. The correspondence of January--March 1903 shows not only a general exchange of ideas, but a concrete circulation of references, tables, diagrams, and proposed calculations in connection with physical chemistry, dissociation, osmotic pressure, diffusion, hydrodynamics, and molecular forces \cite{CPAE5Eng}. In January 1903 Einstein told Besso that he intended first to work on molecular forces in gases and then to undertake comprehensive studies in electron theory. At that moment, he was also studying Richter's \emph{Organic Chemistry}, a standard and recently updated textbook \cite{CPAE5Eng}. This is a significant detail. Einstein was not confining himself to narrowly defined theoretical physics; he was actively acquiring chemical literature relevant to the molecular and physico-chemical problems that would occupy him later.

Besso's reply of 7--11 February 1903 is even more revealing. He sent Einstein a dense set of bibliographical and experimental indications concerning electrolytic dissociation and osmotic phenomena. The names and references he supplied included Battelli and Stefanini, Raoult, Arrhenius, Tammann, Beckmann, Ostwald, Kohlrausch, Roloff, and Sutherland, all figures associated with major contemporary debates in physical chemistry, solution theory, electrolytic dissociation, osmotic pressure, and transport phenomena \cite{CPAE5Eng}. Besso also supplied numerical tables comparing values obtained by different experimental methods, and enclosed two diagrams or sketches concerning incompletely semipermeable membranes and the reduction of osmotic pressure. He explicitly asked Einstein whether certain data could be verified in Bern. The exchange therefore shows a concrete documentary practice: not merely discussion, but the sending of references, tables, diagrams, and requests for local verification.

The presence of William Sutherland in this exchange is especially important. Besso discussed Sutherland's hypothesis on semipermeable membranes and osmotic pressure, while Einstein proposed an experimental criterion for testing it. In his reply of 17 March 1903, Einstein went further. He asked Besso whether he had already calculated the absolute size of ions under the assumption that they were spherical and large enough for the equations of viscous hydrodynamics to apply. He added that, given the known absolute size of electrons, this should be a simple calculation, but that he himself lacked both the necessary literature and the time to carry it out \cite{CPAE5Eng}. Einstein also suggested that diffusion could be used to learn something about neutral salt molecules in solution. These remarks anticipate, in striking form, the cluster of problems that would reappear in Einstein's doctoral dissertation and in his 1905 work on Brownian motion: molecular dimensions, diffusion, osmotic pressure, hydrodynamic resistance, and the determination of microscopic quantities from macroscopic transport phenomena. Yet William Sutherland's related work was not mentioned in those later publications.

The Sutherland case is methodologically important because it shows that Einstein's published citations do not provide a complete map of the literature that entered his working horizon. A topic, author, or problem could be discussed with Besso, become relevant to Einstein's own research, and yet leave no explicit trace in the final published references. This weakens any simple inference from non-citation to non-knowledge. The point is directly relevant to the present article: Einstein's silence about Lorentz or Poincaré in the 1905 relativity paper cannot, by itself, be treated as evidence that their work was absent from his documentary or intellectual horizon.

Einstein's letter of 17 March 1903 is also important for the Bern setting. He wrote that he found it very difficult to put together the material for his paper on molecular forces and complained bitterly about the university in Bern \cite{CPAE5Eng}. In the context of Besso's preceding request that Einstein verify bibliographical or numerical information in Bern, this complaint strongly suggests that Einstein was in fact trying to use the local institutional resources available to him, even if he found them inadequate. From late October 1903 until mid-May 1905, moreover, Einstein lived at Kramgasse 49, about 200 metres from the Hochschulbibliothek at Herrengasse 38 \cite{HentschelGrasshoff2005}. During this period, then, the documentary environment was materially accessible, and Einstein's correspondence shows that he knew how to seek, request, check, and discuss relevant literature when a scientific problem required it.

This evidence is especially important because the subjects discussed with Besso were not peripheral to Einstein's work in the Bern years. They belonged to the same broad domain of molecular physics, diffusion, osmotic pressure, and physical chemistry in which Einstein was then developing his own research. Moreover, Besso's assistance had already taken a technical form before his arrival in Bern: while still in Trieste, about a year earlier, he had carried out calculations for Einstein at Einstein's request. Besso's role was therefore not merely conversational. He functioned, at least in part, as a bibliographical, technical, and intellectual intermediary, helping Einstein orient himself in a literature that was mathematically demanding, experimentally dense, and rapidly developing.

Einstein's own activity as a reviewer for the \textit{Beiblätter zu den Annalen der Physik} points in the same direction. Reviewing required attention to recent publications and placed him within one of the major German-language bibliographical instruments of contemporary physics. This shows that his ordinary scientific practice included the search for relevant literature and the use of bibliographical channels.

His correspondence with Ehrat and Besso, together with his reviewing activity, shows that such practices were part of his scientific life before and during the Bern years. Availability is therefore not equivalent to reading, but neither can Einstein's silence or lack of citation be interpreted against a background of presumed documentary isolation.

\section{Conclusion}

The Saint Louis lecture of September 1904 occupies a distinctive position in the early history of twentieth-century theoretical physics. Historians have long recognized its conceptual importance as one of Poincaré's clearest formulations of the principle of relativity as a general methodological constraint on physical law. What has received less attention is the material history through which this formulation entered the international circulation of scientific ideas. The present article has argued that this material history is not a peripheral bibliographical matter. It bears directly on the documentary landscape within which the emergence of Einstein's 1905 paper on the electrodynamics of moving bodies must be historically situated.

The reconstruction presented here shows that Poincaré's Saint Louis lecture circulated rapidly through several distinct but complementary channels: \textit{La Revue des idées}, the \textit{Bulletin des sciences mathématiques}, the January 1905 issue of \textit{The Monist}, and finally \textit{La valeur de la science}. These were different publication environments, with different readerships, institutional relays, and commercial mechanisms. Together, they made Poincaré's reflections on principles, relativity, Lorentz's theory, time, space, and the future of mechanics available in French and English before Einstein submitted his relativity paper.

The evidence examined in this article also shows that \textit{La valeur de la science} was not a late or inert book object. By the end of April 1905 it was already publicly visible outside France; by 1 May it was being cited and intellectually appropriated in the Belgian press; by 12 May it had reached Göttingen through international bookselling channels; and before the end of May it was advertised in French-speaking Switzerland. These traces establish a chronology of rapid European circulation. They do not prove that Einstein read the book, but they do show that the book belonged to the documentary world of spring 1905.

This conclusion affects the way the Habicht letter should be used. The traditional dating of Einstein's four-paper announcement to 18 or 25 May 1905 should not close the chronological window too quickly. Thursday 1 June 1905 also satisfies the internal evidence of the letter, and Einstein's remark about the expected offprints of the light-quantum paper makes such a date a serious possibility. The Bernese chronology strengthens this point. Einstein's Brownian-motion paper had just been received by the \textit{Annalen}; the Einstein family was in the midst of a documented domestic move; and contemporary administrative records show that Besso changed residence in 1905, from Zeughausgasse 41 to Schwarzenburgstrasse 15, even though the exact date of his physical move cannot presently be fixed. Whether one adopts 25 May or 1 June, the essential point is that Einstein's electrodynamics paper was still being described as a draft at a moment when Poincaré's book had already entered documented European circulation. The date of 1 June, however, fits the reconstructed sequence with particular coherence.

The manuscript testimony concerning the Olympia Academy and Joseph Sauter's recollections further complicates any simple narrative of isolation. Solovine's 1952 letter to Seelig preserves a broader memory of the Academy's readings than the later printed accounts, explicitly including both \textit{La science et l'hypothèse} and \textit{La valeur de la science}. Sauter's testimony, from within the Patent Office milieu, preserves a different but convergent indication: Einstein's relation to Lorentz's formulae at the moment of completion or immediate verification of the relativity paper was not one of simple absence or irrelevance, but of recognized concordance. These testimonies do not establish direct dependence. They show that later printed memory narrowed a denser field of readings, discussions, and retrospective recollections.

The Bern documentary environment points in the same direction. The University Library of Bern held major German, French, Dutch, and international channels of contemporary physics. Einstein's correspondence with Ehrat and Besso, together with his own reviewing activity for the \textit{Beiblätter}, shows that he actively sought recent literature, references, tables, and offprints when a scientific problem required it. The newly reconstructed geography of Einstein's and Besso's Bernese residences gives further concreteness to the remembered conversations on the way home. Availability is therefore not equivalent to reading, but neither can silence or non-citation be interpreted against a background of presumed documentary isolation.

The broader historiographical consequence is that the emergence of special relativity should not be framed as a choice between direct dependence and complete independence. Such an opposition is too crude for the documentary situation reconstructed here. Einstein's originality remains profound: his 1905 paper reorganized the electrodynamics of moving bodies around synchronization, simultaneity, the relativity principle, and the constancy of the velocity of light in a way that differed conceptually from Lorentz's electron theory and from Poincaré's formulation. But originality does not require documentary isolation. A scientific contribution may be conceptually decisive while still belonging to a dense field of available texts, partial formulations, discussions, and problems.

The central result of this article may therefore be stated simply: availability is not influence, but without reconstructing availability the problem of influence is historically ill posed. The evidence presented here makes it difficult to sustain any historiography that begins from the tacit assumption that relevant texts by Poincaré and Lorentz were absent from Einstein's documentary horizon. The task is to situate the making of the 1905 paper within the actual circulation of printed scientific culture and within the concrete documentary conditions of Einstein's Bernese milieu. In that setting, Einstein appears not as an isolated thinker detached from contemporary literature, nor as a passive recipient of prior results, but as an exceptionally powerful reformulator working within a rich and rapidly moving European field of texts, problems, and concepts.

\section*{Acknowledgments}

The author wishes first to thank the two anonymous referees and the editor for their careful reading of the manuscript and for their very important suggestions. Their comments helped to clarify the historical stakes of the article and encouraged the author to broaden, strengthen, and sharpen the argument presented here.

The author also wishes to express his sincere gratitude to the librarians, archivists, and library staff members who provided information on the arrival dates, accession records, institutional stamps, cataloguing details, provenance marks, and historical holdings of \textit{La Revue des idées}, \textit{The Monist}, the \textit{Bulletin des sciences mathématiques}, and \textit{La valeur de la science}. Their generous assistance made possible an essential part of the documentary reconstruction presented in this article. In several cases, their replies allowed the author to distinguish contemporary holdings from later reprints or retrospective acquisitions, and thereby to refine the chronology of circulation before June 1905.

\end{document}